\documentclass[fleqn,twoside]{article}
\usepackage{espcrc2}

\readRCS
$Id: espcrc2.tex,v 1.2 2004/02/24 11:22:11 spepping Exp $
\usepackage{graphicx}
\usepackage[figuresright]{rotating}

\newcommand{\AmS}{{\protect\the\textfont2
  A\kern-.1667em\lower.5ex\hbox{M}\kern-.125emS}}

\title{\vspace*{-10mm}
{\small DESY 08--088; SFB/CPP-08-50
}\\On the tensor reduction of one-loop pentagons and hexagons}
\author{T. Diakonidis\address[DZ]{
Deutsches Elektronen-Synchrotron, DESY,
Platanenallee 6, 15738 Zeuthen, Germany},
        J. Fleischer\addressmark[DZ]\address[B]{
Fakult\"at f\"ur Physik,
Universit\"at Bielefeld,
Universit\"atsstr. 25,
33615 Bielefeld, Germany},
        J. Gluza\address[Kat]{
Institute of Physics, University of Silesia,
Uniwersytecka 4, 40-007 Katowice, Poland},
        K. Kajda\addressmark[Kat],
        T. Riemann\addressmark[DZ],
        J.B. Tausk\addressmark[DZ]
      }

\runtitle{}
\runauthor{}

\begin{document}

\begin{abstract}
We perform analytical reductions of one-loop tensor integrals with 5
and 6 legs to scalar master integrals. They are based on the use
of recurrence relations connecting integrals in different space-time
dimensions. The reductions are expressed in a compact form in terms of
signed minors, and have been implemented in a mathematica package called
{\tt hexagon.m}. We present several numerical examples.
\vspace{1pc}
\end{abstract}

\maketitle

\section{Introduction}

Recent years have seen the emergence of first results for loop corrections to
massive $2\to4$
scattering processes~\cite{Denner:2005es,Boudjema:2005rk,Pozzorini}.
One of the challenges posed by such processes is the need to compute
one-loop tensor integrals with up to 6 legs. In this contribution,
we concentrate on integrals with 5 and 6 legs.

In 4-dimensional space-time, a linear relation must exist between
the loop momentum and the external momenta of a pentagon integral.
Similarly, if the external momenta of a hexagon are 4-dimensional,
they must be linearly dependent. This enables one to reduce
hexagons to pentagons, and pentagons to
boxes~\cite{Melrose:1965kb,van Neerven:1983vr,van Oldenborgh:1989wn}.
These considerations have been extended to dimensionally regularized
one-loop tensor integrals~\cite{Bern:1992em,Binoth:1999sp}.
Generally, in the reduction of tensor integrals one encounters
inverse Gram determinants, which can vanish at exceptional
phase space points and be a source of instabilities in a numerical program.
In the case of 4-dimensional pentagons, it is possible to avoid
the leading inverse Gram determinant~\cite{Denner:2002ii}. Other schemes
for the reduction of multileg one-loop tensor integrals have
been discussed in
refs.~\cite{Fleischer:1999hq,Duplancic:2003tv,Binoth:2005ff,Denner:2005nn}.

Here, we use the methods of ref.~\cite{Fleischer:1999hq}.
Our present goal is to provide compact analytic formulas for the
complete reduction of tensor pentagons and hexagons to scalar
master integrals, which are free of leading inverse Gram determinants.
We describe an implementation of these formulas in a mathematica package
{\tt hexagon.m}.

\section{Notations}

We consider one-loop, $N$-point tensor integrals of rank $R$
in $d$-dimensional space-time,
\begin{equation}
\label{eq:JNR}
J^{(N)}_{\mu_1\ldots\mu_R} \left(d;\nu_1,\ldots,\nu_N\right) =
\int \frac{d^d k}{i \pi^{d/2}}
\frac{k_{\mu_1}\ldots k_{\mu_R}}{D_1^{\nu_1}\ldots D_N^{\nu_N}}
\end{equation}
with propagator denominators\footnote{
In order to be consistent with the conventions of
ref.~\cite{Fleischer:1999hq},
we have changed the sign of $q_j$ as compared
with the original definition of ref.~\cite{Davydychev:1991va}.}
\begin{equation}
D_j = (k-q_j)^2 - m_j^2 + i \epsilon \, .
\end{equation}
Following Davydychev~\cite{Davydychev:1991va}, we decompose
these tensor integrals into a basis of symmetric tensors
constructed from metric tensors $g$ and the momenta $q_j$
\begin{eqnarray}
\label{eq:dav}
\lefteqn{
J^{(N)}_{\mu_1\ldots\mu_R} \left(d;\nu_1,\ldots,\nu_N\right) =
{(-1)}^R
\sum_{
\lambda,\kappa_1,\ldots,\kappa_N
}
{(-\frac{1}{2})}^{\lambda}
} && \nonumber \\ && \hspace{-1em}
\left\{[g]^\lambda [q_1]^{\kappa_1}
 \ldots [q_N]^{\kappa_N}\right\}_{\mu_1\ldots\mu_R}
{(\nu_1)}_{\kappa_1} \ldots {(\nu_N)}_{\kappa_N}
\nonumber \\ && \hspace{-1em}
J^{(N)} \left(d + 2(R-\lambda) ;
\nu_1 + \kappa_1,\ldots,\nu_N + \kappa_N\right)
\end{eqnarray}
where 
$
{(\nu)}_{\kappa} = \frac{\Gamma(\nu+\kappa)}{\Gamma{(\nu)}}
$
are Pochhammer symbols and the sum runs over non-negative
integers such that
$ 2 \lambda + \kappa_1 + \ldots + \kappa_N = R $.
The next step is to use recurrence relations to reduce the scalar
coefficients $J^{(N)}$ appearing in the decomposition to a set of master
integrals.

It is useful to introduce a notation for certain determinants
that occur in the recurrence relations and their solutions. First,
the determinant of an $(N+1)\times(N+1)$ matrix, known as the modified
Cayley determinant~\cite{Melrose:1965kb},
\begin{equation}
\label{eq:smN}
()_N ~\equiv~  \left|
\begin{array}{ccccc}
  0 & 1       & 1       &\ldots & 1      \\
  1 & Y_{11}  & Y_{12}  &\ldots & Y_{1N} \\
  1 & Y_{12}  & Y_{22}  &\ldots & Y_{2N} \\
  \vdots  & \vdots  & \vdots  &\ddots & \vdots \\
  1 & Y_{1N}  & Y_{2N}  &\ldots & Y_{NN}
\end{array}
\right| \, ,
\end{equation}
with coefficients
\begin{equation}
Y_{ij}=-(q_i-q_j)^2+m_i^2+m_j^2 \, , \quad (i,j = 1 \ldots N) \, .
\end{equation}
Although the masses of the propagators appear in the coefficients
$Y_{ij}$, the determinant $()_N$ does not depend on them and it is
actually proportional to the Gram determinant of the external
momenta of the $N$-point function in eq.~(\ref{eq:JNR}).
All other determinants we need are signed minors of $()_N$, constructed
by deleting $m$ rows and $m$ columns
from $()_N$, and multiplying with a sign factor. They will be denoted
by
\begin{eqnarray}
\lefteqn{
\left(
\begin{array}{cccc}
  j_1 & j_2 & \cdots & j_m\\
  k_1 & k_2 & \cdots & k_m\\
\end{array}
\right)_N
~\equiv~ {(-1)}^{\sum_l (j_l + k_l)}
} && \nonumber \\ && 
\hspace{-1.5em}
\mbox{sgn}_{\{j\}} \, \mbox{sgn}_{\{k\}} \,
 \left| 
\begin{array}{c}
\mbox{rows $j_1\cdots j_m$ deleted}\\
\mbox{columns $k_1\cdots k_m$ deleted}\\
\end{array}
 \right| \, ,
\end{eqnarray}
where $\mbox{sgn}_{\{j\}}$ and $\mbox{sgn}_{\{k\}}$ are the signs of
permutations that sort the deleted rows $j_1\cdots j_m$ and columns
$k_1\cdots k_m$ into ascending order.

Combining integration by parts identities with relations
connecting integrals in different space-time
dimensions~\cite{Tarasov:1996br}, one obtains the
following basic recurrence relations~\cite{Fleischer:1999hq}:
\begin{eqnarray}
\label{eq:RR1}
\lefteqn{
\left( \right)_N
 \nu_j{\bf j^+} J^{(N)}(d+2) =
} && \nonumber \\ &&
\left[  - {j \choose 0}_N +\sum_{k=1}^{n} {j \choose k}_N
 {\bf k^-} \right] J^{(N)}(d) \, ,
\end{eqnarray}
\begin{eqnarray}
\label{eq:RR2}
\lefteqn{
  (d-\sum_{i=1}^{n}\nu_i+1) \left(  \right)_N J^{(N)}(d+2)
  =
} && \nonumber \\ &&
  \left[ {0 \choose 0}_N
 - \sum_{k=1}^n {0 \choose k}_N {\bf k^-} \right] J^{(N)}(d) \, ,
\end{eqnarray}
\begin{eqnarray}
\lefteqn{
{0\choose 0}_N \nu_j {\bf j^+} J^{(N)}(d) =
  \sum^{n}_{k=1} {0j\choose 0k}_N 
} && \nonumber \\ && \times
\left[ d - \sum_{i=1}^{n} \nu_i( {\bf k^-} {\bf i^+}+1)
             \right] J^{(N)}(d) \, .  
\end{eqnarray}
where the operator ${\bf j^{\pm}}$ acts by shifting
the index $\nu_j$ by $\pm 1$.

\section{Pentagons}
A detailed discussion of the second rank pentagon is given in
Ref.~\cite{Fleischer:2007ph}.
In this section, we will consider a third rank tensor integral
with indices $\nu_1 = \ldots = \nu_5 = 1$, which we write as
$I_{5}^{\mu\, \nu\, \lambda}$.
We assume here and in the next section,
that the loop momentum $k$ has been shifted in such a way that $q_N=0$.
Applying eq.~(\ref{eq:dav}) gives integrals in space-time dimensions
$d+4$ and $d+6$ and with increased indices. They are reduced back to
the generic dimension $d=4-2\epsilon$ by the recurrence relations 
in eqs.~(\ref{eq:RR1})-(\ref{eq:RR2}). This involves a division
by a Gram determinant $()_N$ at each step. The leading Gram determinant,
$()_5$, can be avoided if one is only interested in contractions
of the tensor integral with 4-dimensional
objects~\cite{Denner:2002ii}. This is achieved by using the following
decomposition of the metric tensor,
\begin{equation}
g^{\mu\nu} ~=~
 2\, \sum_{i,j=1}^{N-1} \frac{{i\choose j}_N}{{\left( \right)}_N}
 \, q_i^{\mu} q_j^{\nu}
 \, + \, g_{\perp}^{\mu\nu} \, ,
\end{equation}
and dropping terms proportional to $g_{\perp}^{\mu\nu}$. After
further simplifications we obtain:
\begin{equation}
\label{eq:I5mnl}
I_{5}^{\mu\, \nu\, \lambda} =
 \sum_{i,j,k=1}^{4} \, q_i^{\mu}\, q_j^{\nu} \, q_k^{\lambda} E_{ijk}
+\sum_{k=1}^4 g^{[\mu \nu} q_k^{\lambda]} E_{00k} \, ,
\nonumber \\
\end{equation}
where
\begin{equation}
g^{[\mu \nu} q_k^{\lambda]}
 =       g^{\mu \nu}      \, q_k^{\lambda}
 \, + \, g^{\mu \lambda}  \, q_k^{\nu}
 \, + \, g^{\nu \lambda}  \, q_k^{\mu} \, ,
\end{equation}
with scalar coefficients defined by
\begin{eqnarray}
\label{eq:Eijk}
\lefteqn{
E_{ijk} =
   \sum_{s=1}^{5} S_{ijk}^{4,s} I_4^s
}
&& \nonumber \\ &&
 {} + \sum_{s,t=1}^{5} S_{ijk}^{3,st} I_3^{st}
    + \sum_{s,t,u=1}^{5} S_{ijk}^{2,stu} I_2^{stu} \, ,
\end{eqnarray}
with
\begin{eqnarray}
\lefteqn{
S_{ijk}^{4,s}=\frac{1}{3 {0\choose 0}_5  {s\choose s}_5^2}
\times } &&
 \nonumber\\ & &
 \left\{-{0s\choose 0k}_5 \left[{0s\choose is}_5 {0s\choose js}_5 +
{is\choose js}_5 {0s\choose 0s}_5 \right]
 \right.
\nonumber \\ && \left.
{} + {0s\choose 0s}_5 \left[
{0i\choose sk}_5 {0s\choose js}_5+
{0j\choose sk}_5 {0s\choose is}_5 \right]
 \right\} \nonumber\\ &&
 {} + (i \leftrightarrow k) + (j \leftrightarrow k) \, ,
\end{eqnarray}
\begin{eqnarray}
\lefteqn{
S_{ijk}^{3,st}=\frac{1}{3 {0\choose 0}_5  {s\choose s}_5^2 } 
\left\{ {0s\choose 0k}_5   
\left[ {ts\choose is}_5 {0s\choose js}_5
\right. \right. } && \nonumber \\ &&
\left.
 {} + {is\choose js}_5 {ts\choose 0s}_5 
    + \frac{{s\choose s}_5 {0st\choose ist}_5}{{st\choose st}_5}
 {ts\choose js}_5\right]
 \nonumber\\ &&
 {} -\left[{0i\choose sk}_5{0s\choose js}_5
    +{0j\choose sk}_5{0s\choose is}_5\right]{ts\choose 0s}_5
 \nonumber\\ &&
 {} -\left[{0i\choose sk}_5{ts\choose js}_5
   +{0j\choose sk}_5{ts\choose is}_5\right]
 \nonumber\\ && \left.
 \times \frac{{s\choose s}_5 {0st\choose 0st}_5}{ 2 {st\choose st}_5}
 \right\}
 + (i \leftrightarrow k) + (j \leftrightarrow k) \, ,
\end{eqnarray}
\begin{eqnarray}
\lefteqn{
S_{ijk}^{2,stu} =
 -\frac{1}{3 {0\choose 0}_5 {s\choose s}_5 {st\choose st}_5}
\times } && 
 \nonumber\\ &&
 \left\{ {0s\choose 0k}_5  {ts\choose js}_5 {ust\choose ist}_5
 - \frac{1}{2}
 \left[{0j\choose sk}_5 {ust\choose ist}_5
 \right.\right. \nonumber\\ && \left.\left.
 {} + {0i\choose sk}_5 {ust\choose jst}_5\right] {ts\choose 0s}_5
 \right\}
 \nonumber\\ &&
 {} + (i \leftrightarrow k) + (j \leftrightarrow k) \, ,
\end{eqnarray}
and
\begin{eqnarray} 
\label{eq:E00j}
\lefteqn{
 E_{00j} = \frac{1}{6 {0\choose 0}_5}
\left\{
 - \sum_{s=1}^5~~ \frac{1}{{s\choose s}_5^2}~
\right. } &&
\nonumber \\ && \times
\left[~3 {s\choose 0}_5 {0s\choose js}_5
       {} - {s\choose j}_5 {0s\choose 0s}_5 \right]
{0s\choose 0s}_5 I_{4}^{s}
\nonumber \\ &&
 {}+ 
 \sum_{s,t=1}^5~~  \frac{1}{{s\choose s}_5^2}~ 
\nonumber \\ && \times
\left[~3 {s\choose 0}_5 {0s\choose js}_5
 - {s\choose j}_5 \frac{{ts\choose 0s}_5^2 }{{st\choose st}_5}
 ~\right]{ts\choose 0s}_5 I_{3  }^{st}
\nonumber \\ && \left.
 {}- 
 \sum_{s,t,u=1}^5
{s\choose j}_5 
 \frac{{ust\choose 0st}_5}
{{s\choose s}_5 {st\choose st}_5}~~{ts \choose 0s}_5 I_{2  }^{stu}
\right\} \, .
\end{eqnarray} 
The decomposition in eq.~(\ref{eq:I5mnl}) is equivalent to the one
found in ref.~\cite{Denner:2002ii}, where the coefficients $E_{ijk}$
and $E_{00j}$ are expressed in terms of tensor 4-point functions. Here,
instead, they are completely reduced to a basis of scalar master integrals
consisting of boxes $I_{4}^{s}$, vertices $I_{3  }^{st}$, and 2-point
functions $I_{2  }^{stu}$ obtained by removing lines $s$, $s$ and $t$,
or $s$, $t$ and $u$ from the pentagon.

\section{Hexagons}
If the external momenta of a hexagon are 4-dimensional,
their Gram determinant vanishes: $ \left( \right)_6 = 0 $, and
a linear relation between the propagators $D_j$ exists:
\begin{equation}
1 ~=~
\sum_{j=1}^6 \frac{{0\choose j}_6}{{0\choose 0}_6} D_j \, .
\end{equation}
With this relation, any hexagon integral can trivially be
reduced to pentagons. For example, for the scalar hexagon,
one obtains the well-known result~\cite{Melrose:1965kb}:
\begin{equation}
I_6=\sum_{r=1}^6 \frac{{0\choose r}_6}{{0\choose 0}_6} I_5^r \, ,
\end{equation}
where the scalar pentagon $I_5^r$ on the right hand side is
obtained by removing line $r$ from the hexagon $I_6$.
In the same way, tensor hexagons of rank $R$ can be reduced to tensor
pentagons of rank $R$. However, it was noticed in
ref.~\cite{Fleischer:1999hq} that a reduction directly to tensor
pentagons of rank $R-1$ is also possible:
\begin{equation}
\label{eq:I6tensor}
I_6^{\mu_1\ldots\mu_R} ~=~
 \sum_{r=1}^6 v_r^{\mu_1} I_5^{\mu_2\ldots\mu_R\,,r} \, ,
\end{equation}
where
\begin{equation}
v_r^{\mu} ~\equiv~ 
 - \frac{1}{{0\choose 0}_6} \sum_{i=1}^{5} {0i\choose 0r}_6 q_i^{\mu} \, .
\end{equation}
A more general proof of this property was given in ref.~\cite{Denner:2005nn}.
By substituting our reduction formulas for tensor pentagons into
eq.~(\ref{eq:I6tensor}), we can immediately express tensor
hexagons in terms of scalar master integrals.

\section{Numerical results}

\begin{figure}[!htb]
        \centerline{ \includegraphics[scale=0.75]{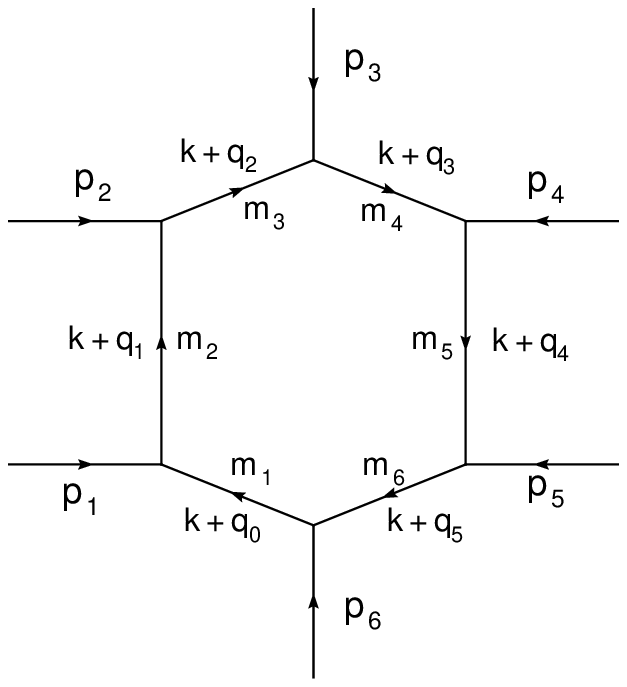} }
	\centerline{ \includegraphics[scale=0.75]{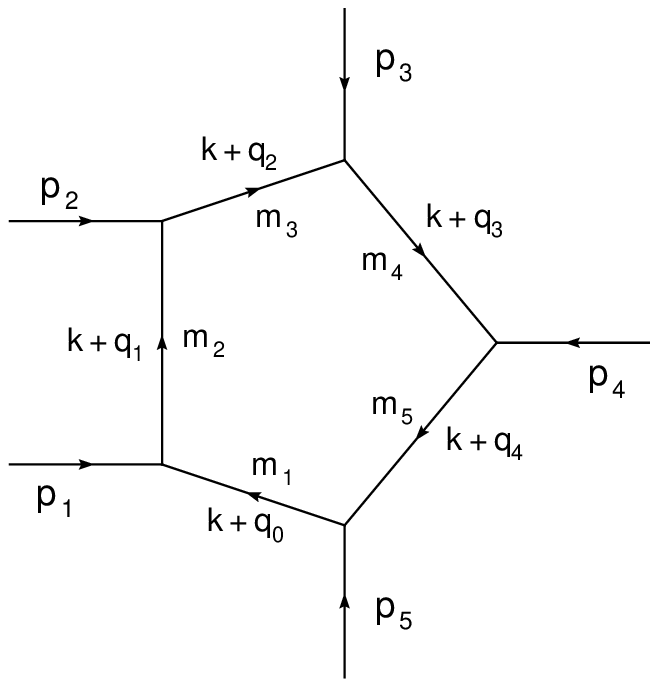} }
	\caption{Momentum flow used in {\tt hexagon.m} for six and five point
                 diagrams.}
   \label{fig:6pt5ptfig}
\end{figure}

The method of reducing six and five point Feynman integrals presented
in this paper was implemented in a {\tt MATHEMATICA} package called
\verb#hexagon.m#~\cite{hexagon}.
It allows to reduce six point integrals. Additionally,
as hexagons are connected with pentagons, the package also allows
one to reduce pentagons.
The present implementation includes:
\begin{itemize}
	\item six point functions up to rank four
	\item five point functions up to rank three
\end{itemize}
These tensor ranks are sufficient to get results for e.g.
NNLO contributions to Bhabha scattering.
The kinematics used in the package\footnote{
The masses and momenta are numbered according to the conventions
of {\tt LoopTools}~\cite{Hahn:1998yk}, which are slightly different from
those used in the previous sections and in ref.~\cite{Fleischer:1999hq}.}
is presented in Fig.~\ref{fig:6pt5ptfig}.\\
Before using \verb#hexagon.m#, the package must be loaded in
a {\tt MATHEMATICA} environment by executing:
\begin{verbatim}
	<<hexagon.m
\end{verbatim}
The package is able to output the full result for a six or five point
tensor integral,
a specific coefficient, or a list of all coefficients for a given rank.
For a more detailed description, see Table~\ref{tab:1}.
\begin{table*}[htb]
\caption{Functions used in the package.}
\label{tab:1}
\begin{tabular}{@{}ll|ll}
\hline
\multicolumn{2}{@{}l|}{Six point functions} 
&
\multicolumn{2}{@{}|l}{Five point functions} 
\\
\hline
\verb#RedF0#
&
scalar 6pt integral
&
\verb#RedE0#
&
scalar 5pt integral
\\
\verb#RedF1#
&
vector 6pt integral
&
\verb#RedE1#
&
vector 5pt integral
\\
\verb#RedF2#
&
rank two 6pt tensor integral
&
\verb#RedE2#
&
rank two 5pt tensor integral
\\
\verb#RedF3#
&
rank three 6pt tensor integral
&
\verb#RedE3#
&
rank three 5pt tensor integral
\\
\verb#RedF4#
&
rank four 6pt tensor integral
&
&
\\
\hline
\verb#RedFcoef#
&
coefficient of given 6pt
&
\verb#RedEcoef#
&
coefficient of given 5pt
\\
\verb#RedFget#
&
all coefficients of given 6pt
&
\verb#RedEget#
&
all coefficients of given 5pt
\\
\hline
\multicolumn{4}{@{}l}{
The basic functions have the following arguments, here
$s_{ij}=(p_i+p_j)^2$, $s_{ijk}=(p_i+p_j+p_k)^2$:}
\end{tabular}
\verb#RedF0[#$p_1^2,\dots,p_6^2,s_{12},s_{23},s_{34},s_{45},s_{56},s_{16},s_{123},s_{234},s_{345},m_1^2,\dots,m_6^2$\verb#]#\\
\verb#RedE0[#$p_1^2,\dots,p_5^2,s_{12},s_{23},s_{34},s_{45},s_{15},m_1^2,\dots,m_5^2$\verb#]#
\end{table*}
\begin{table*}[htb]
\caption{Numerical cross-checks.}
\label{tab:2}
\begin{tabular}{@{}l}
\hline
1) Comparison with {\tt AMBRE \& MB.m} :
$p_1^{\mu} p_1^{\nu} p_1^{\lambda} E_{\mu\nu\lambda}$
\\\hline
Point:
\\
$p_1^2=p_2^2=p_3^2=p_5^2=1$, $p_4^2=0$, 
$m_1^2=m_3^2=0$, $m_2^2=m_4^2=m_5^2=1$, 
\\
$s_{12}=-3$, $s_{23}=-6$, $s_{34}=-5$, $s_{45}=-7$, $s_{15}=-2$
\\\hline
\verb#In: RedE3[#
$p_1^2,\dots,p_5^2,s_{12},s_{23},s_{34},s_{45},s_{15},m_1^2,\dots,m_5^2$
\verb#]/.{D4->D0,C3->C0,B2->B0}#
\\
\verb#Out: 0.218741# 
\\\hline
2) Comparison with {\tt Sector Decomposition} :
$F_0$
\\\hline
Point:
\\
$p_1^2=p_2^2=p_3^2=p_4^2=p_5^2=p_6^2=-1$, 
$m_1^2=m_2^2=m_3^2=m_4^2=m_5^2=m_6^2=1$, 
\\
$s_{12}=s_{23}=s_{34}=s_{45}=s_{56}=s_{16}=s_{123}=s_{234}=-1$, $s_{345}=-5/2$
\\\hline
\verb#In: RedF0[#
$p_1^2,\dots,p_6^2,s_{12},s_{23},s_{34},s_{45},s_{56},s_{16},s_{123},s_{234},s_{345},m_1^2,\dots,m_6^2$
\verb#]/.{D4->D0}#
\\
\verb#Out: 0.013526# 
\\\hline
3) Comparison with {\tt LoopTools} :
$E_0$, $E_1$, $E_2$, $E_3$, $E_4$, $E_{34}$, $E_{123}$, $E_{002}$
\\\hline
Point:
\\
$p_1^2=p_2^2=0$, $p_3^2=p_5^2=49/256$, $p_4^2=9/100$, 
$m_1^2=m_2^2=m_3^2=49/256$, $m_4^2=m_5^2=81/1600$, 
\\
$s_{12}=4$, $s_{23}=-1/5$, $s_{34}=1/5$, $s_{45}=3/10$, $s_{15}=-1/2$
\\\hline
\verb#In: RedE0[#
$p_1^2,\dots,p_5^2,s_{12},s_{23},s_{34},s_{45},s_{15},m_1^2,\dots,m_5^2$
\verb#]/.D4->D0#
\\
\verb#Out: 41.3403 - 45.9721*I#
\\
\verb#In: RedEget[rank1#
$,p_1^2,\dots,p_5^2,s_{12},s_{23},s_{34},s_{45},s_{15},m_1^2,\dots,m_5^2$
\verb#]/.D4->D0#
\\
\verb#Out: ee1 =-2.38605 + 5.27599*I, ee2 =-5.80875 + 0.597891*I,# 
\\
\verb#     ee3 =-14.4931 + 20.8149*I, ee4 =-11.3362 + 18.1593*I#
\\
\verb#In: RedEcoef[ee34#
$,p_1^2,\dots,p_5^2,s_{12},s_{23},s_{34},s_{45},s_{15},m_1^2,\dots,m_5^2$
\verb#]/.{D4->D0,C3->C0}#
\\
\verb#Out: 7.1964 + 3.10115*I#
\\
\verb#In: RedEcoef[ee123#
$,p_1^2,\dots,p_5^2,s_{12},s_{23},s_{34},s_{45},s_{15},m_1^2,\dots,m_5^2$
\verb#]/.{D4->D0,C3->C0,B2->B0}#
\\
\verb#Out:-0.149527 - 0.31059*I# 
\\
\verb#In: RedEcoef[ee002#
$,p_1^2,\dots,p_5^2,s_{12},s_{23},s_{34},s_{45},s_{15},m_1^2,\dots,m_5^2$
\verb#]/.{D4->D0,C3->C0,B2->B0}#
\\
\verb#Out: 0.154517 - 0.387727*I#
\\\hline
\end{tabular}

\end{table*}
\verb#hexagon.m# is able to generate both analytic and numerical results,
depending on the user's input. It provides coefficients of Lorentz-covariant
tensors, and works in a basis of $g^{\mu \nu}$ and internal momenta (chords)
$q_i$, 
\begin{equation}
	q_0=0, \quad q_n=\sum_{i=1}^{n}p_i,
\end{equation}
see also Fig.~\ref{fig:6pt5ptfig}. In terms of these
coefficients, the tensor decomposition of pentagons $E$ and hexagons
$F$ reads:
\begin{eqnarray}
	E^\mu &=& \sum_{i=1}^{4}q_i^\mu E_i,
\nonumber\\
	E^{\mu \nu} &=& \sum_{i,j=1}^{4}q_i^\mu q_i^\nu E_{ij}+
		g^{\mu \nu} E_{00},
\nonumber 
\end{eqnarray}\vspace*{-10mm}\begin{eqnarray}
	E^{\mu \nu \lambda} &=& 
		\sum_{i,j,k=1}^{4}q_i^\mu q_i^\nu q_k^\lambda E_{ijk}+
		\sum_{i=1}^{4} 
		g^{[\mu \nu} q_i^{\lambda ]} E_{00i},
\nonumber\\
	F^\mu &=& \sum_{i=1}^{5}q_i^\mu F_i,
\nonumber\\
	F^{\mu \nu} &=& \sum_{i,j=1}^{5}q_i^\mu q_i^\nu F_{ij},
\nonumber\\
   F^{\mu \nu \lambda} &=& 
		\sum_{i,j,k=1}^{5}q_i^\mu q_i^\nu q_k^\lambda F_{ijk}+
		\sum_{i=1}^{5} 
		g^{\mu \nu} q_i^{\lambda} F_{00i},	
\nonumber\\
	F^{\mu \nu \lambda \rho} &=&		
	\sum_{i,j,k,l=1}^{5}q_i^{\mu} q_i^{\nu} q_k^{\lambda} q_l^{\rho} F_{ijkl}
\nonumber\\	
		&+&
		\sum_{i,j=1}^{5} 
		q_i^{\mu} q_j^{[ \nu} g^{\lambda \rho ] } F_{00ij}.
\end{eqnarray}
For the purpose of checking the correctness of the reduction procedures
implemented in the package, we have made both internal and external checks.
Internal checks were used for tensor integrals, and consisted mainly in
writing a scalar product of internal and external momenta in terms
of lower rank tensor integrals, which had been checked before. External
checks were made with use of:  {\tt LoopTools}~\cite{Hahn:1998yk}
(five point integrals); {\tt AMBRE} \cite{Gluza:2007rt} with {\tt MB.m}
\cite{Czakon:2005rk} (five point integrals); {\tt Sector Decomposition}
\cite{Bogner:2007cr} (five and six point scalar integrals).

We present some of the cross-checks in Table~\ref{tab:2}. In all these
checks, we use {\tt LoopTools} to calculate the finite parts of the scalar
four, three and two point functions which appear after the reduction by
{\tt hexagon.m}. We note that, in general, the functions defined directly
in {\tt LoopTools} may not be sufficient to cover the whole kinematic
phase space obtained from reduction of six point functions. In such cases
our reduction package must be supplemented with additional libraries.

In the first example in Table~\ref{tab:2}, we use {\tt AMBRE} and
{\tt MB.m} to check the decomposition in the Euclidean kinematic region,
including tensor structures. We show the result from {\tt hexagon.m}
for a contracted tensor of rank three. At the level typical for Monte
Carlo calculations, it is in agreement with the result from Mellin-Barnes
integration: $0.218885$.

The second example comes from ref.~\cite{Binoth:2002xh}, (Table 2, region I).
Here, we use the package~\cite{Bogner:2007cr} and show for the case of
the scalar six point function agreement with the calculation using
the sector decomposition method (typically five digits accuracy).
In the third example, we have extended the scalar case given
in~\cite{Binoth:2002xh} (Table 1, region III). Also the tensorial results
agree directly with {\tt LoopTools}.

Finally, for six point tensor integrals, we have performed comparisons with a
Fortran implementation of our reduction formulas by two of us (TD and BT),
and also with an independent code by P. Uwer~\cite{PU}. Some sample
results, for the randomly chosen phase space point given in
table~\ref{tab:PSP}, are shown in table~\ref{tab:hexagonResults}.
\begin{table*}[htb]
\caption{Phase space point used in the tensor hexagons}
\label{tab:PSP}
\begin{eqnarray*}
p_1&=&(  0.21774554E+03,
\quad    0,
\quad    0,
\quad    0.21774554E+03)
\\
p_2&=&(  0.21774554E+03,
\quad    0,
\quad    0,
\quad   -0.21774554E+03)
\\
p_3&=&( -0.20369415E+03,
\quad   -0.47579512E+02,
\quad    0.42126823E+02,
\quad    0.84097181E+02)
\\
p_4&=&( -0.20907237E+03,
\quad    0.55215961E+02,
\quad   -0.46692034E+02,
\quad   -0.90010087E+02)
\\
p_5&=&( -0.68463308E+01,
\quad    0.53063195E+01,
\quad    0.29698267E+01,
\quad   -0.31456871E+01)
\\
p_6&=&( -0.15878244E+02,
\quad   -0.12942769E+02,
\quad    0.15953850E+01,
\quad    0.90585932E+01)
\\ &&
m_1 = 110.0, \;
m_2 = 120.0, \;
m_3 = 130.0, \;
m_4 = 140.0, \;
m_5 = 150.0, \;
m_6 = 160.0
\end{eqnarray*}
\end{table*}

\begin{table}[htb]
\caption{Results for scalar, vector, and 2nd-rank six point functions
for the phase space point of Table~\ref{tab:PSP}.}
\label{tab:hexagonResults}
\centering
\vspace{5pt}
\begin{tabular}{|l l|c|c|}
\hline
&&\multicolumn{2}{|c|}{\bf{RESULTS}} \\
\hline
& &\multicolumn{1}{|c|}{REAL} &\multicolumn{1}{|c|}{IM}  \\ \hline
& &\multicolumn{2}{|c|}{$F_0$} \\ \hline
& &-0.223393E-18 &  -0.396728E-19\\\hline
$\mu$ & &\multicolumn{2}{|c|}{$F^{\mu}$} \\ \hline
0&     &0.192487E-17 &   0.972635E-17 \\ 
1&    &-0.363320E-17 &  -0.11940E-17  \\ 
2&     &0.365514E-17 &   0.106928E-17 \\ 
3&     &0.239793E-16 &   0.341928E-17 \\ \hline
$\mu$&$\nu$&\multicolumn{2}{|c|}{$F^{\mu\nu}$} \\ \hline
0&0&   0.599459E-14  & -0.114601E-14 \\ 
0&1&    0.323869E-15  &  0.423754E-15\\ 
0&2&   -0.294252E-15  & -0.375481E-15\\ 
0&3&   -0.255450E-14  & -0.195640E-14\\ 
1&1&   -0.164562E-14  & -0.993796E-16\\ 
1&2&    0.920944E-16  &  0.706487E-17\\ 
1&3&    0.347694E-15  & -0.127190E-16\\ 
2&2&   -0.163339E-14  & -0.994148E-16\\ 
2&3&   -0.341773E-15  &  0.818678E-17\\ 
3&3&   -0.413909E-14  &  0.670676E-15\\ \hline
\end{tabular}
\end{table}

\section{Summary}
We have described an analytical reduction of one-loop tensor integrals
with 5 or 6 legs down to scalar master integrals, and shown explicitly the
result for the tensor pentagon of rank three. Formulas for other cases
and detailed derivations will be presented elsewhere. The reduction
formulas have been implemented in a mathematica package {\tt hexagon.m},
which will be made publicly available.

\section*{Acknowledgments}
We thank P.~Uwer for discussions and for his kind assistance
in cross checking our results for the tensor hexagons.
Work supported by Sonderforschungsbereich/Transregio
SFB/TRR 9 of DFG ``Computergest\"utzte Theoretische Teilchenphysik"
and by the European Community's Marie-Curie Research Training Networks
MRTN-CT-2006-035505 ``HEPTOOLS'' and MRTN-CT-2006-035482 ``FLAVIAnet''.

\end{document}